%% file: main.tex
\documentclass[10pt,conference,letterpaper]{IEEEtran}

\usepackage{cite}
\usepackage{amsmath,amssymb,amsfonts}
\usepackage{algorithmic}
\usepackage{graphicx}
\usepackage{textcomp}
\usepackage[table]{xcolor}
\usepackage{xcolor}
\usepackage{dblfloatfix}
\usepackage{circledsteps}

\def\BibTeX{{\rm B\kern-.05em{\sc i\kern-.025em b}\kern-.08em
    T\kern-.1667em\lower.7ex\hbox{E}\kern-.125emX}}
    
\begin{document}

\title{The Effect of the Russian-Ukrainian Conflict from the Perspective of Internet eXchanges
}

\author{\IEEEauthorblockN{1\textsuperscript{st} Cristian Trusin}
\IEEEauthorblockA{\
\textit{University of Twente}\\
The Netherlands \\
c.trusin@student.utwente.nl}
\and
\IEEEauthorblockN{2\textsuperscript{nd} Leandro Bertholdo}
\IEEEauthorblockA{
\textit{University of Twente}\\
The Netherlands \\
l.m.bertholdo@utwente.nl}
\and
\IEEEauthorblockN{3\textsuperscript{rd} José Jair Santanna}
\IEEEauthorblockA{
\textit{Northwave and University of Twente}\\
The Netherlands \\
jair.santanna@northwave.nl}
}

\maketitle

\begin{abstract}
In 2022 the Russian invasion of Ukraine began. It is known that Ukraine faced outages because of the damage to their infrastructure. It is also known that Russia was boycotted by the international community. However the impact on the telecommunications of the two countries remains unknown. In this paper we quantified the degree to which the Internet was affected in both countries by analyzing routing tables from five large Internet Exchange Points (IXPs). IXPs provide a central point of interconnection where internet traffic can be freely exchanged between Autonomous Systems (ASes). This centrality makes IXPs a good vantage point for analyzing changes in the Internet infrastructure. With data collected before and after the start of the conflict we observed considerable damage to the Ukrainian Internet network with numerous outages and minimal damage to the Russian network. An average of 11.12\% of Ukrainian ASes were unreachable at each IXP. We identified the biggest outages and the events responsible for them. This paper highlights resilience issues during conflicts to the network and management community, and serves as a basis for future more in-depth research.
\end{abstract}


\input{Sections/Introduction}

\begin{figure*}[b]
\centering
\includegraphics[width = \textwidth]{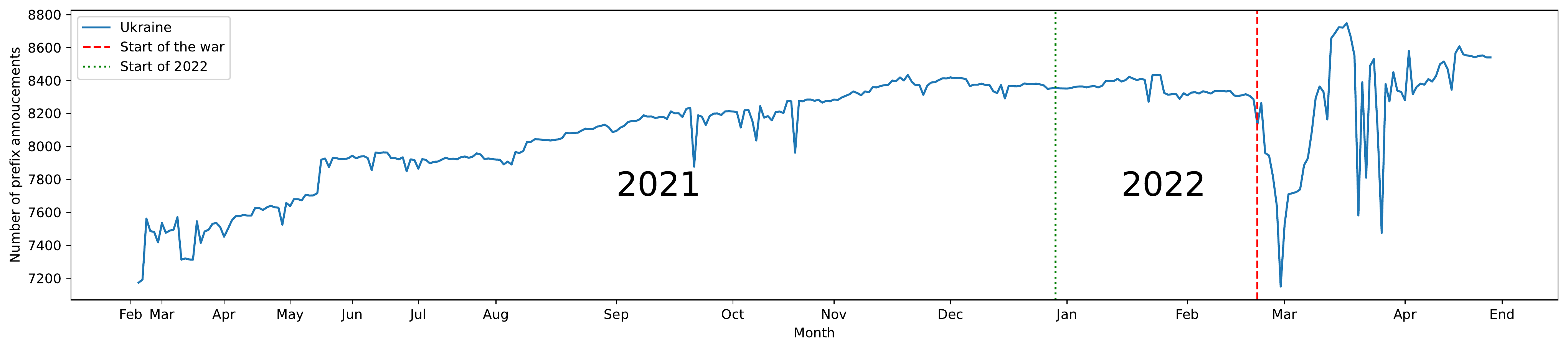}
\caption{Number of Ukrainian Prefix announcements before the start of the war.}
\label{fig1}
\end{figure*}

\input{Sections/Background}

\input{Sections/Methodology}

\begin{figure*}[t]
\centering
\includegraphics[width = \textwidth]{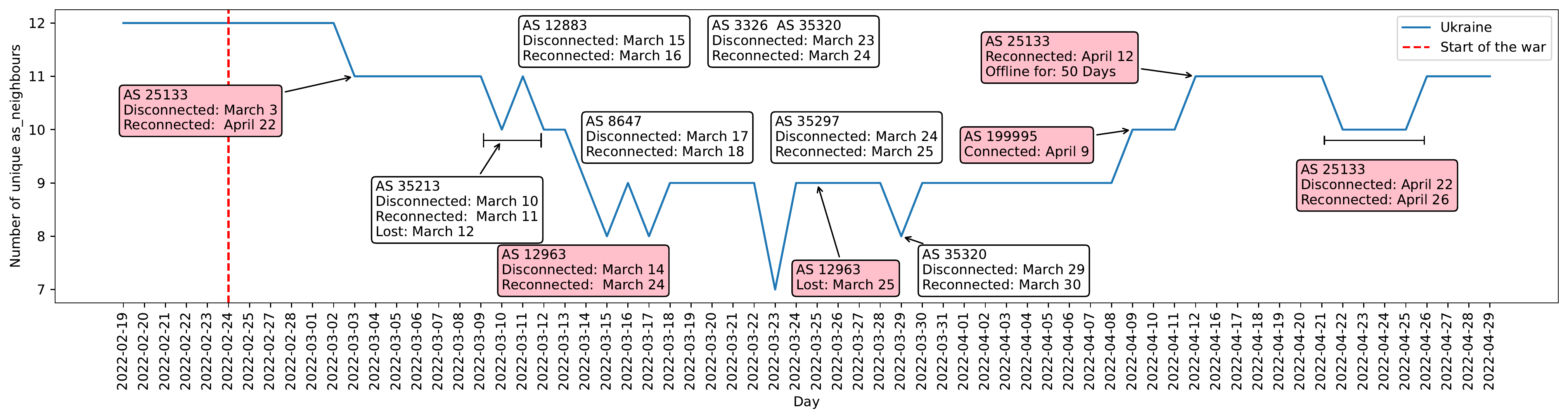}
\caption{Number of Ukrainian \textit{as\_neighbors} at AMSIX.}
\label{fig2}
\end{figure*}

\begin{figure*}[b]
\centering
\includegraphics[width = \textwidth]{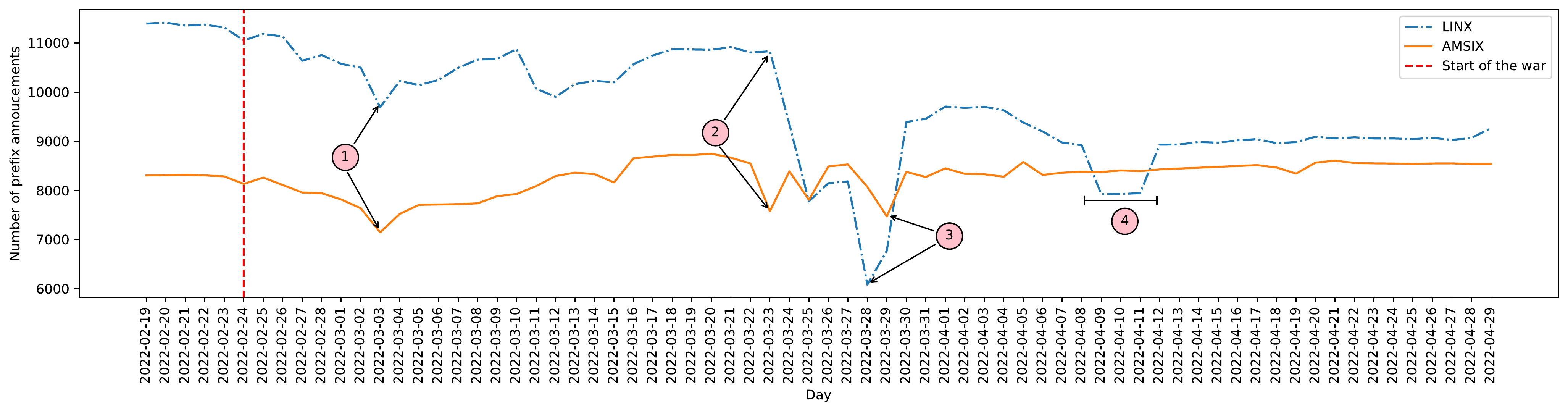}
\caption{Number of Ukrainian Prefix announcements at AMSIX and LINX.}
\label{fig3}
\end{figure*}

\begin{figure*}[t]
\centering
\includegraphics[width = \textwidth]{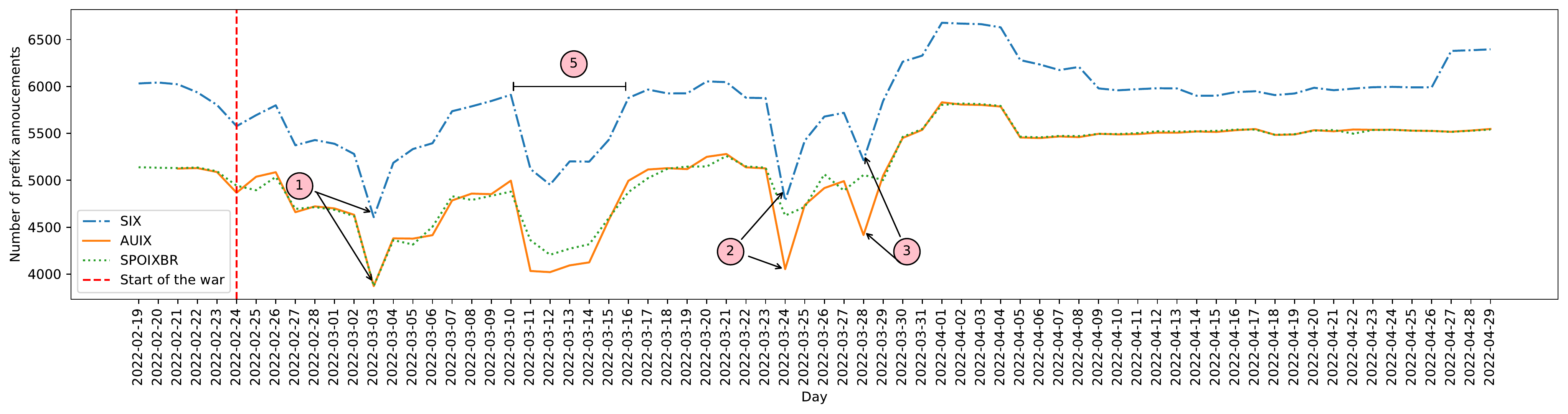}
\caption{Number of Ukrainian Prefix announcements at SIX, AUIX and SPOIXBR.}
\label{fig4}
\end{figure*}

\begin{figure*}[b]
\centering
\includegraphics[width = \textwidth]{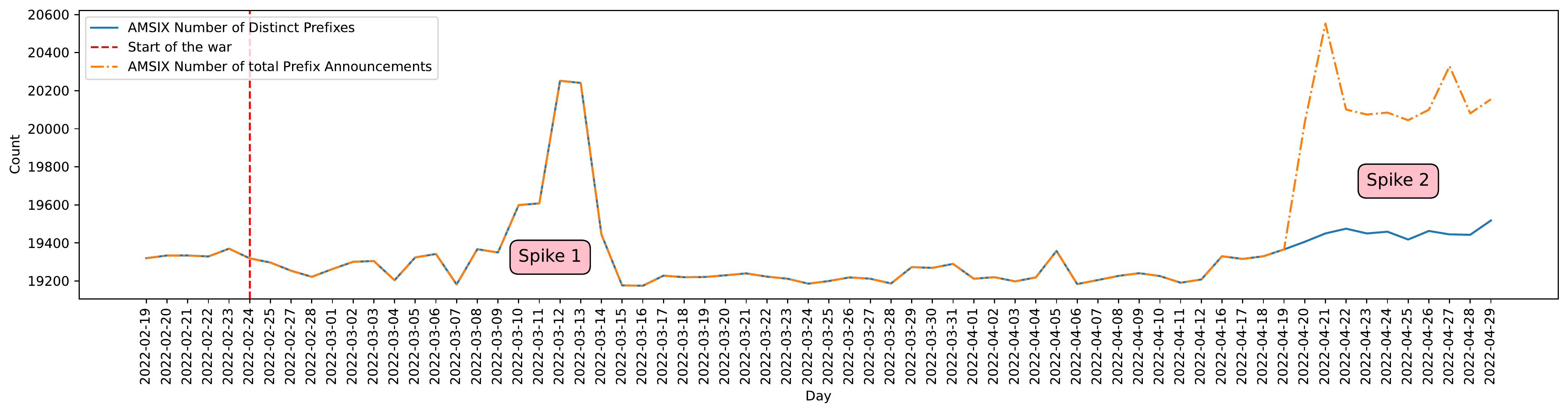}
\caption{Number of Russian Connections at AMSIX.}
\label{fig5}
\end{figure*}

\begin{figure*}[t]
\centering
\includegraphics[width = \textwidth]{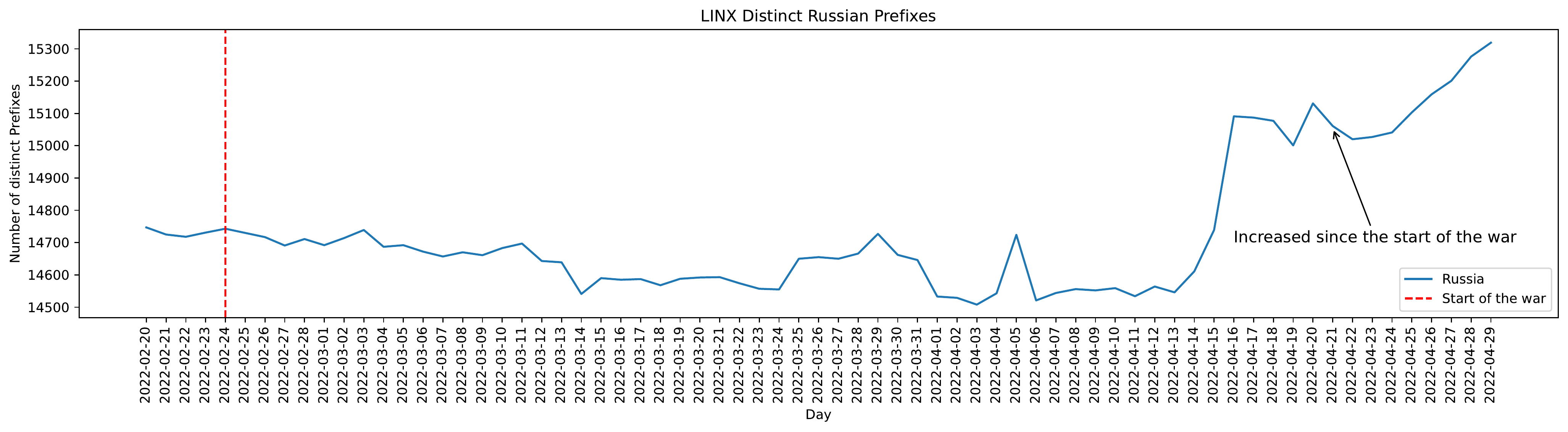}
\caption{Number of Distinct Russian Prefixes at LINX.}
\label{fig6}
\end{figure*}

\begin{figure*}[b]
\centering
\includegraphics[width = \textwidth]{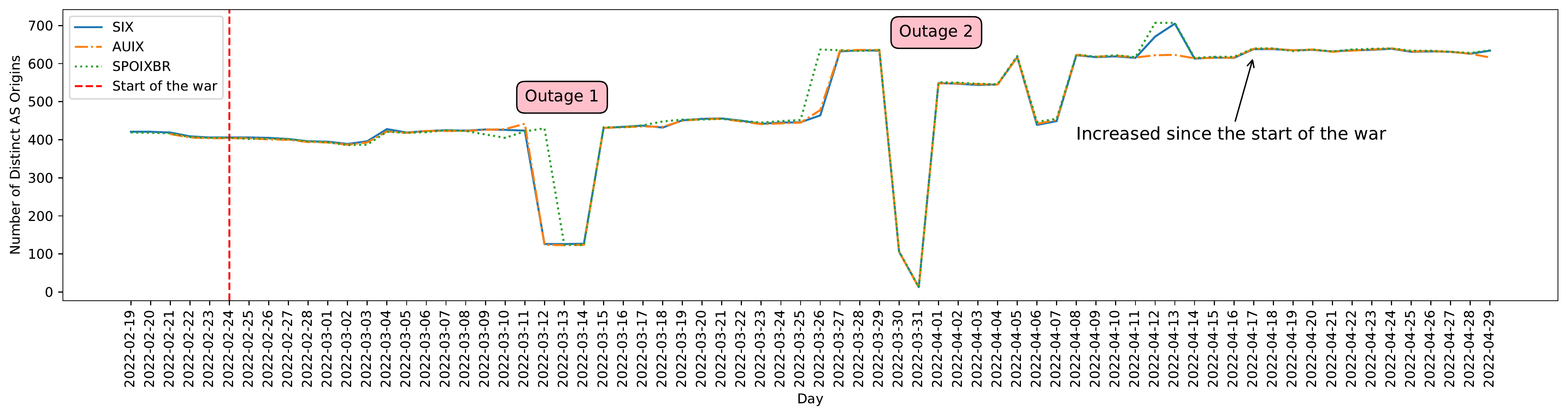}
\caption{Number of Distinct Russian \textit{as\_origins} at SIX, AUIX and SPOIXBR.}
\label{fig7}
\end{figure*}

\input{Sections/Results}
\input{Sections/Conclusions_and_Future_Work}

\section*{Acknowledgment}
This work is funded by the Netherlands Organisation for Scientific Research,
PAADDoS project (628.001.029), 
and CONCORDIA, the Cybersecurity Competence Network supported by
the European Union\textquotesingle s Horizon 2020 research and innovation
programme under grant agreement No 830927.
We would like to thank Suzan Bayhan for her continued support for this paper.

\bibliographystyle{IEEEtran}
\bibliography{references}

\end{document}

%% file: Sections/Introduction.tex
\section{Introduction}
On the 24\textsuperscript{th} of February 2022 Russia invaded Ukraine. Ukraine has been experiencing blackouts as parts of its infrastructure have been cut off after bombings \cite{1}. Russia has been boycotted by the international community and businesses. Sanctions were registered since 24\textsuperscript{th} of February 2022 and are still ongoing \cite{2}. Major transit providers like Cogent and Lumen, stopped selling Internet access to Internet Service Providers (ISPs) inside Russia \cite{2,3}. Furthermore, content providers like TikTok and Netflix have ceased their services in the country \cite{4}. These events have had an impact on the Internet in both countries, some services are unreachable or are forced to be routed through a slower path, but it remains unclear how big this impact is.

In this paper we investigate some of this impact using data from Internet Exchange Points (IXPs). IXPs are a crucial element of today’s internet \cite{5}, they provide direct peering opportunities between Autonomous Systems (ASes). Direct peering is a voluntary interconnection of two separate networks with the purpose of directly exchanging the traffic between the users of those networks, it improves latency and decreases interconnection costs. Since IXPs connect many ASes, they are a rich source for networking research \cite{5}, and are suitable to analyze changes in the network, thus they can help us quantify the effects of the war on the Internet in Ukraine and Russia.

Our methodology to assess this impact is based on Border Gateway Protocol (BGP) data from the routing tables provided by five large IXPs: AMSIX, LINX, SIX, AUIX, and SPOIXBR. We only investigated open peering data from these IXPs and we analyze the period between 19\textsuperscript{th} of February 2022 (before Russian invasion) and 29\textsuperscript{th} of April 2022 (more than two months after the invasion). Our goal is to quantify the effect that the war had on the Russian and Ukrainian networks from the perspective of the IXPs.

To achieve this goal, we have defined the following research questions (RQs) as a basis for our research:
\begin{itemize}
    \item RQ1: How many Russian and Ukrainian ASes became unreachable?
    \item RQ2: For how long were these ASes unreachable?
    \item RQ3: What were the biggest outages and what events are responsible for them?
\end{itemize}

This paper is structured as follows: In Section 2 we discuss the necessary background information and the related work. Section 3 will talk about the methodology we used to answer the Research Questions. Section 4 will discuss the results of each research question and their implications. In Section 5 we conclude our work and discuss possibilities for future work.

%% file: Sections/Background.tex
\section{Background}
The Internet is a network of networks, it can be broken into smaller networks called Autonomous Systems (ASes). Each AS is a group of routers that belong to the same organization. For example, to ISPs, universities, tech companies, governments etc. Each AS wishing to connect to the Internet and exchange routing information must have a valid AS Number (ASN) assigned by the Internet Assigned Numbers Authority.

ASes communicate with each other using the Border Gateway Protocol (BGP). BGP provides a standardized way of sharing routing information between ASes, it is the glue that connects all the Autonomous Systems together. At its essence, BGP is trying to connect the whole Internet together by letting ASes share their routing data. Then from this routing data, it selects the best path to a destination based on an attribute called \textit{as\_path}. The \textit{as\_path} is a sequence of ASNs that a packet will follow to reach the destination. BGP chooses the shortest \textit{as\_paths}. The details of this process are not important for this paper, but more information can be found in 
\cite{6}. 

For two ASes to interconnect, they need to establish a peering session. One way of doing this is to connect to an IXP, which is a central point of interconnection for ASes, and form an open peering agreement where they can freely exchange traffic with other members of the IXP. This is achieved by a Route Server.  A Route Server is a construct in an IXP which has information about the routes offered by other member ASes. By connecting to a Route Server, you can establish connection to multiple ASes by using a single BGP session, this enables one-to-many peerings instead of traditional one-to-one peerings \cite{7}, which improves scalability and makes routing easier. By connecting to only 10 IXPs you can reach 56\% of all ASes on the Internet \cite{8}, which makes them a good source of insight to answer our research questions. A good high-level introduction to IXPs can be found in Gerson’s et al. \cite{9}, a more detailed dive into IXPs can be found in \cite{5}.

%% file: Sections/Methodology.tex
\section{Methodology}
In this section we will explain how we performed our research. We will start by explaining what this dataset is and how it was collected, and then dive into our implementation. We quantified the outages and changes in the network by performing Data Analysis on selected BGP attributes from our collected dataset. The code for this implementation as well as a Database required for it is provided here \cite{10}.
\subsection{Datasets and IXP Selection}
From 13\textsuperscript{th} of February 2021 to 29\textsuperscript{th} of April 2022, the routing tables from AMSIX, LINX, SIX, AUIX, SPOIXBR have been collected daily through a Looking Glass, as part of the research conducted by Bertholdo et al. \cite{8}. A Looking Glass allows an IXP to share information from the Route Server. The routing tables represent only open peering data.

The 5 IXPs listed above were chosen due to their size and representativeness \cite{8}, \cite{11}. AMSIX (Amsterdam Internet Exchange) and LINX (London Internet Exchange) are in Europe, close to the conflict. LINX is particularly interesting since they reported sanctions against Russia \cite{12}. AMSIX is the second biggest IXP in Europe after DE-CIX (Deutscher Commercial Internet Exchange). We found some gaps in the DE-CIX dataset during war days, so it was not possible to include it in our analysis. We also included other IXPs outside Europe to verify if problems were observed on other continents. SIX (Seattle Internet Exchange) and AUIX (Australia Internet Exchange) were used to represent North America and Australia in this research. And finally for South America we had data from SPOIXBR (S$\tilde{a}$o Paulo Internet Exchange in Brazil).

The gathered routing tables are in a .csv format and represent BGP data. We have a routing table for every single day and for each IXP. So, for any given day we have 5 routing tables to analyze since we have 5 IXPs. We will focus on the \textit{as\_origin}, \textit{as\_neighbor} and \textit{as\_path} attributes of the routing tables. \textit{as\_origin} denotes the AS that can be reached from this IXP, it is an ASN. If this AS cannot be reached anymore, its \textit{as\_origin} will disappear from the routing table. One \textit{as\_origin} can route traffic to multiple prefixes. \textit{as\_neighbor} from the perspective of an IXP denotes the ASN of the AS that is directly connected to the IXP. \textit{as\_path} denotes all the ASes the \textit{as\_origin} had to transit through to reach the IXP. \textit{as\_path} is a list of ASNs, the last entry in the list is always the \textit{as\_origin}. We will use these attributes to analyze if and how routes to Ukrainian and Russian ASes have changed.

We know for a fact that certain ASes experienced outages because of the war \cite{1}, \cite{4}. Moreover, the routing paths to some Russian ASes have changed \cite{3}. To quantify these, we needed a baseline. We analyzed several days of data before the war as can be seen in Figure 1, and concluded that except for 2 outages in September and October 2021, the data showed little variation, so we chose 19\textsuperscript{th} of February 2022 to serve as our baseline. The last datapoint we have is on 29\textsuperscript{th} of April (End label in Figure 1).

\subsection{Implementation}
The first phase of the research was identifying which ASNs belong to Russia and Ukraine. We did this by creating a database where we aggregated data from Regional Internet Registries (RIRs). This data offers information about the current allocations and assignments of Internet Number Resources (IPs and ASNs). We used the data from all 5 RIRs: AFRINIC, APNIC, ARIN, LANIC and RIPE NCC. We filtered it to contain ASN assignments only, and for each ASN we kept the country to which it belongs. The data from each RIR was later aggregated together, resulting in a Database that can identify the originating country of a given ASN. From now on we will refer to this Database as ASNDB. ASNDB allows us to filter our routing tables and isolate Ukrainian and Russian routes.

\subsubsection{Counting the number of unreachable ASes (RQ1)}
To answer RQ1 we needed to count the number of ASes which became unreachable at any point after the start of the war. We joined ASNDB with our routing tables, resulting in a table where for each route we also have its originating country code as a new column. By filtering on this new column, we were able to find the Ukrainian and Russian \textit{as\_origins} for each IXP. By analyzing our baseline and looking at our last datapoint, we identified which and how many \textit{as\_origins} became unreachable after the start of the war for each IXP. This was achieved by selecting all distinct prewar \textit{as\_origins} and searching for those \textit{as\_origins} in the last routing table we have collected. If an \textit{as\_origin} was not found, it was pronounced unreachable. It could be that a temporary outage was happening during the collection of our last datapoint, making some \textit{as\_origins} temporarily unreachable. To avoid drawing false conclusions we also checked if they were unreachable during the 3 days before the collection of our last datapoint. The results did not show any significant differences.

\subsubsection{Counting the number of days ASes were unreachable (RQ2)}
Due to the sheer amount of ASes and variation in the number of days they were offline for, we could not just count for how long each AS was offline, since this data would be hard to visualize. We decided to focus on the \textit{as\_neighbors} instead. An \textit{as\_neighbor}, in the context of an IXP, is the ASN of the AS which is directly connected to the IXP. This decision was made because if one \textit{as\_neighbor} were to disconnect, all the prefixes it provided routing information for will also be disconnected, thus making them unreachable for this IXP and the member ASes of this IXP. This is indicative of an outage at the said AS, since the consequences of disconnecting are undesirable. If an ISP were to disconnect from an IXP, all its clients would lose access to fast routes connecting them to the global internet. Bearing this in mind, we counted the number of distinct \textit{as\_neighbors} for each day at every IXP.

\subsubsection{Identifying the outages and the responsible events (RQ3)}
To answer RQ3 we needed to find when the outages happened. We could not use the plot from RQ2 for this purpose, since it only offers a high-level view of the situation. It could be that the \textit{as\_neighbor} has not disconnected from the IXP, but it has fewer prefixes to route to because of the damage to the infrastructure.

Instead, we calculated the number of Ukrainian and Russian prefix announcements, alongside the number of distinct Ukrainian and Russian \textit{as\_origins} for every day, for each IXP. This was achieved by joining the ASNDB with the routing tables and filtering on routes originating from the two countries. We plotted this data to see where we can find dips in the plot (outages). Having found a dip we would look at the news  for that day alongside the report from NetBlocks \cite{1} to find events that are responsible for the outage.

%% file: Sections/Results.tex
\section{Results and Discussion}
In this section we show and discuss our findings. Since there is little overlap between the situation of Ukrainian and Russian Networks, we split it into 2 subsections: one discussing our results regarding Ukraine and another regarding Russia. The plots used show data since 19\textsuperscript{th} of February, since we chose this date as our baseline of the prewar situation. The datapoints before this day showed little variation as discussed in Section III Part A, so they were cut off from the resulting plot.

\subsection{Mapping outages in Ukraine}
The Ukrainian network took considerable damage since the start of the war. Table I provides information about the number of unreachable Ukrainian ASes at each IXP. On average every single IXP lost 11.12\% of Ukrainian ASes since the war started.  The European IXPs were the most affected of all.

\begin{table}[htbp]
\caption{Number of unreachable Ukrainian \textit{\textit{as\_origins}}}
\begin{center}
\begin{tabular}{|>{\columncolor[gray]{0.8}}c|c|>{\columncolor[gray]{0.8}}c|c|}
\hline
\textbf{IXP} & \textbf{Total ASes} & \textbf{Lost ASes} & \textbf{\% Lost} \\
\hline
AUIX & 1016 & 87 & 8.5\%  \\
\hline
LINX & 1335 & 254 & 19.0\% \\
\hline
AMSIX & 1571 & 164 & 10.4\% \\
\hline
SPOIXBR & 1021 & 92 & 9.0\% \\
\hline
SIX & 1096 & 96 & 8.7\% \\
\hline
\end{tabular}
\label{tab1}
\end{center}
\end{table}

\subsubsection{Europe}
AMSIX experienced the most variation in available Ukrainian connections. Each \textit{as\_neighbor} provides routes to many \textit{\textit{as\_origins}}, losing one is significant. At AMSIX, a Ukrainian \textit{as\_neighbor} provides routes to an average of 400 prefixes which is around 409.600 IP Addresses. As we can see in the historical graph in Figure 2, most of these ASes experienced an outage for a day, however some (AS 25133 – McLaut-Invest) were unreachable for 50 days, and some never reconnected (AS 12963 – Volz). A peculiar finding is Omegatelecom (AS 199995) which is a new member of AMSIX coming from Ukraine. It connected on April 9th and is responsible for providing routes to 243 distinct Ukrainian prefixes and 90.264 IP addresses. LINX has not experienced any significant loss of \textit{as\_neighbor}s. The only disconnection happened on 28th until 29th of March during Event \Circled{3} in which LINX lost its only Ukrainian \textit{as\_neighbor} which provided routes to many Ukrainian prefixes. This is explained by a cyberattack on Ukrtelecom’s core infrastructure as reported by BBC and NetBlocks \cite{14, 17} and is also responsible for the decrease in prefix announcements during that time. 

As can be seen in Figure 3, there are significant variations in the number of prefix announcements at AMSIX and LINX since the start of the war. Event \Circled{1} shows a significant loss of connectivity, around 1000 less prefixes were announced that day. This can be attributed to a major blackout in the region of Sumy, which is reported as the largest disruption to Telcom services since the start of the war \cite{15}. As a result of bombing at the local thermal power plant, the region of Sumy experienced a large blackout which also affected Internet connectivity. Connectivity started to improve over the course of the next days, and 10 days later new prefix announcements were observed, even more than before the war.

Other major disruptions can be seen at Event \Circled{2} and Event \Circled{3}. We could not identify a concrete event responsible for the former, but the latter can be attributed to the cyberattack on Ukrtelecom’s core infrastructure mentioned above \cite{14, 17}. Overall Ukrainian prefixes at AMSIX seem to have rebound to prewar levels since April. 

However, LINX does not show such a rebound. It sustained the biggest Ukrainian losses, registering a total of 254 out of 1335 ASes. 19\% of all Ukrainian networks were disconnected from this IXP. Figure 3 shows that LINX has been experiencing a gradual decline in Ukrainian connections since the beginning of the war. Besides experiencing the same outages as AMSIX, LINX also had a loss of connectivity during Event \Circled{4}, an emergency was reported by the operator WNET which impacted Ukraine’s international connectivity during that period \cite{16}.

\subsubsection{Other continents}
Figure 4 shows the data analyzed for SIX (USA), AUIX (Australia) and SPOIXBR (Brazil). These IXPs follow a similar pattern to Europe, with a significant loss of connectivity at Event \Circled{1}, due to the Sumy Blackout \cite{15}, unidentified outage at Event \Circled{2} and cyberattack at Event \Circled{3} \cite{17}. Another dip in connections can be seen during Event \Circled{5}. This can possibly be explained by a global Internet Disruption in mid-March, when the number of Internet outages increased by 22\% \cite{18}. None of these IXPs have Ukrainian \textit{as\_neighbors}, thus none of them were lost during the war.

\subsubsection{Observations and Discussion}
Ukraine has sustained damage to its networks since the start of the war, an average of 11.12\% of Ukrainian ASes are now unreachable. We have found 5 outages: Event \Circled{1} - The Sumy Oblast Blackout, Event \Circled{2} - Unidentified Outage, Event \Circled{3} - Ukrtelecom Cyberattack, Event \Circled{4} - WNET International Connectivity emergency, Event \Circled{5} - March Global Internet Disruption. The first three of these can be observed both in and outside Europe. Event \Circled{4} is exclusive to LINX, and Event \Circled{5} is exclusive to IXPs outside Europe. Except for LINX, the number of Ukrainian connections at IXPs has started recovering and the number of outages has decreased significantly since April.

LINX had the biggest losses, we observed a decrease of 19\% in the number of reachable Ukrainian \textit{as\_origins}.  Judging by the fact that Event \Circled{3} marks the biggest outage for LINX with a loss of 2000 prefix announcements, and that LINX temporarily lost its only Ukrainian \textit{as\_neighbor} during that event, we can assume that these losses are attributed to the disconnection of that neighbor. This \textit{as\_neighbor} provided routes to most Ukrainian prefixes at LINX and because of the war, it likely lost connection to a lot of ASes, so it could no longer provide routes to those ASes for LINX.

AMSIX has 12 Ukrainian \textit{as\_neighbors} which provided more redundancy. If one neighbor lost connection to some ASes, another one might still have a way to reach them, which explains why AMSIX restored most connections. LINX did not lose all Ukrainian routes due to the disconnection of the said \textit{as\_neighbor} because neighboring countries kept providing alternative routes to some Ukrainian ASes.

These observations invite a discussion about the resiliency of the Ukrainian network. While damaged, it remains functional. This can be attributed to the presence of many redundancies in many layers of the network \cite{RIPE}. Ukraine has 19 local IXPs and they share the market almost evenly. This provides alternative routes in case of outages i.e. redundancy. Moreover, Ukrainian ISPs are very decentralized, 55\% of the ISPs serve less than 1\% of the market \cite{RIPE}, meaning that the Ukrainian population is connected through a variety of ISPs without a single point of failure. However we have observed that energy and cyberattacks are a weak-point of the Ukrainian network. As seen in Event \Circled{1}, damaging energy sources has severe effects on the connectivity of the country since the local ISPs and infrastructure need energy to operate. Event \Circled{3} showed the damage a cyberattack can inflict, however judging by the lack of further outages caused by cyberattacks, it is safe to say that the Ukrainian network learned to mitigate them. Perhaps it could be attributed to the help that Ukraine got from companies like Microsoft and ESET \cite{eset}.

\subsection{The effects of the boycott on Russia's connectivity to IXPs}
Russia was heavily boycotted for their invasion by most countries in the world \cite{1}. There have been threats that Russia will be cut off from the global Internet. Moreover, major Internet Transit Providers like Cogent and Lumen have stopped offering Russia their services \cite{2}, \cite{3}.  In this section we analyze how the Russian network was affected at the IXPs.

Table II provides information about the number of Russian ASes that lost access to the IXP since the start the of the war. An average of 10.94\% ASes were lost. However, new ASes have connected during that same period, thus increasing the total number of Russian \textit{as\_origins} and prefixes since the start of the war.

\begin{table}[htbp]
\caption{Number of unreachable Russian \textit{\textit{as\_origins}}}
\begin{center}
\begin{tabular}{|>{\columncolor[gray]{0.8}}c|c|>{\columncolor[gray]{0.8}}c|c|}
\hline
\textbf{IXP} & \textbf{Total ASes} & \textbf{Lost ASes} & \textbf{\% Lost} \\
\hline
AUIX & 3749 & 117 & 3.1\%  \\
\hline
LINX & 2886 & 109 & 3.7\% \\
\hline
AMSIX & 421 & 62 & 14.7\% \\
\hline
SPOIXBR & 415 & 78 & 18.7\% \\
\hline
SIX & 419 & 61 & 14.5\% \\
\hline
\end{tabular}
\label{tab2}
\end{center}
\end{table}

\subsubsection{Europe}
As can be seen in Figure 5, AMSIX shows no decrease in Russian prefixes pnnouncements and the number of distinct prefixes. A sudden increase in prefix Announcements was observed for 6 days (Fig. 5 Spike 1). We analyzed the number of distinct Russian prefixes at AMSIX during this period and found an increase of about 800 prefixes. Another such spike can be seen (Fig. 5 Spike 2), however this time it was only followed by an increase of 200 distinct prefixes, and it is still ongoing. There was no variation in the number of Russian \textit{as\_neighbors} at AMSIX. 

LINX tells a similar story to AMSIX, there was no decrease in the number of Russian routes and \textit{as\_neighbors}. As can be seen in Figure 6, an increase of distinct prefixes was observed from 15\textsuperscript{th} of April. These findings are peculiar since LINX was one of the few IXPs that announced sanctions against Russia, and one would expect the number of Russian connections to decrease. After further investigation we found that LINX only disconnected two Russian ASes: Rostelecom (AS 12389) and Megafon (AS 31133) \cite{12}. These ASes are Russia’s biggest ISPs, but apparently, they had little to no traffic at LINX and were planning on abandoning LINX services themselves \cite{19}.

\subsubsection{Other Continents}
Figure 7 shows the data for SIX, AUIX and SPOIXBR. They all show that Russia experienced an outage on 12\textsuperscript{th} until 14\textsuperscript{th} of March (Fig. 7 Outage 1), which is consistent with the outage observed at these IXPs for Ukraine (Fig 4 Event \Circled{5}) in the same period. It is also explained by the mid-March global Internet Disruptions \cite{18}. Another serious outage was observed on 31\textsuperscript{st} of March (Fig. 7 Outage 2). SIX and AUIX had no Russian \textit{as\_neighbors} prewar, and the situation remained unchanged. SPOIXBR had one Russian \textit{as\_neighbor} and besides a small 1-day outage on 2\textsuperscript{nd} of March it remained stable. These IXPs also show an increase in Russian prefixes in April, as well as an increase in distinct \textit{as\_origins} since the start of the war. 

\subsubsection{Observations and Discussion}
Russia has not sustained much damage to their network, contrary to the rumors circling in press that it will be disconnected from the global Internet. We found a decrease of 10.94\% in the number of distinct \textit{as\_origins} since the start of the war. However, as we observed in the plots the number of distinct \textit{as\_origins} has increased after the war, making this loss less significant since the total number of Russian ASes has grown since the start of the war. The loss of the old ASes could just signify a change in Russia’s internal Internet network infrastructure. Moreover, the number of Russian prefixes has increased, suggesting that the sanctions applied to Russia did not have a major impact on their network. Our findings align with the ones seen in the report by ThousandEyes \cite{3}.

Major transit providers Cogent and Lumen have stopped their services to Russia. This did not have any significant effects since we did not register a drop in overseas Russian prefixes at SIX, AUIX and SPOIXBR. This can be explained by the numerous alternative transit providers that Russia can still use \cite{20}. The reason behind the ban was limiting Russia’s capacity to perform cyberattacks, according to CNN \cite{21}, and not disconnecting Russia from the Internet.

IXPs have not ceased providing services to Russia in any meaningful way. LINX has stopped providing services to the two biggest Russian ISPs: Rostelecom and Megafon. However, these ISPs were barely using LINX’s services already, and LINX continues to advertise prefixes which contain those ISPs in their \textit{as\_path}. According to the statement provided by LINX to their customers, these measures were not designed to block Russia from the Internet but rather to target individuals which are affiliated with the current Russian government and responsible for the war \cite{12}. Some of them are in direct ownership or benefit from Rostelecom and Megafon. Europe imposed sanctions against these persons \cite{22} and LINX was merely complying with them.

%% file: Sections/Conclusions_and_Future_Work.tex
\section{Conclusions and Future Work}
In this paper we quantified the effect of the conflict between Russia and Ukraine from the perspective of five large IXPs. We found how many Ukrainian and Russian ASes were lost, for how long, and we identified the biggest outages and the events responsible for them.

Our results show that each IXP has lost connection to an average of 11.12\% of Ukrainian ASes. We identified 5 big outages which are responsible for this loss. LINX lost most Ukrainian connections and has not yet restored all of them. Other IXPs started recovering since April and the number of outages has decreased. The Ukrainian network shows resilience, and while clearly damaged remains functional due to the presence of many redundancies.

We did not find any substantial damage, or loss of connectivity to the Russian network. Contrary to the sanctions imposed, the Russian network remains reachable and shows signs of growth.

As this research focused on a limited set of IXPs, future work could analyze data from other IXPs. In particular DE-CIX, since it is the largest one in Europe and from the limited data we had, we noticed that it had a lot of Russian and Ukrainian connections, so it might provide some new insights. It would also be worthwhile to look at the data we analyzed from a different angle, by counting the number of unique reachable IP addresses, thus leading to more fine-grained results. We would also like to analyze changes in the \textit{as\_paths}, and investigate the impact the war had on other countries, in particular neighboring ex-soviet republics, since they may rely on routes provided by Ukraine and Russia.